\RequirePackage{lineno}

\documentclass[prl,superscriptaddress,twocolumn,showpacs,amsmath,amssymb]{revtex4}
\usepackage{graphicx}  
\usepackage{dcolumn}   
\usepackage{bm}        
\usepackage{amssymb}   
\usepackage{multirow}

\begin{document}

\newcommand{\dzero}     {D0}
\newcommand{\ttbar}     {\mbox{$t\bar{t}$}}
\newcommand{\bbbar}     {\mbox{$b\bar{b}$}}
\newcommand{\ccbar}     {\mbox{$c\bar{c}$}}
\newcommand{\herwig}    {{\sc{herwig}}}
\newcommand{\pythia}    {{\sc{pythia}~}}
\newcommand{\fewz}    {{\sc{fewz}~}}
\newcommand{\madgraph}    {{\sc{madgraph}~}}
\newcommand{\vbp}    {{\sc{vbp}~}}
\newcommand{\resbos}    {{\sc{resbos}~}}
\newcommand{\powheg}    {{\sc{powheg}~}}
\newcommand{\dyrad}    {{\sc{dyrad}~}}
\newcommand{\vecbos}    {{\sc{vecbos}}}
\newcommand{\alpgen}    {{\sc{alpgen}}}
\newcommand{\qq}        {\sc{qq}}
\newcommand{\evtgen}    {{\sc{evtgen}}}
\newcommand{\tauola}    {{\sc{tauola}}}
\newcommand{\geant}     {{\sc{geant}}}
\newcommand{\metcal}    {\mbox{$\not\!\!E_{Tcal}$}}
\newcommand{\met}       {\mbox{$\not\!\!E_T$}}
\newcommand{\pt}        {$p_T$}
\newcommand{\sigmatt}   {\ensuremath{\sigma_{t\bar{t}}}}
\newcommand{\mtop}      {$m_{\rm top}$}
\newcommand{\rsigma}    {\ensuremath{R_{\sigma}}}

\newcommand{\feynhiggs} {{\sc{FeynHiggs}}}
\newcommand{\cpsuperh} {{\sc{CPsuperH}}}

\newcommand{\brh}{\ensuremath{{ B}(t\rightarrow H^+ b)}} 

\newcommand{\pythian}    {{\sc{pythia}}}
\newcommand{\fewzn}    {{\sc{fewz}}}
\newcommand{\madgraphn}    {{\sc{madgraph}}}
\newcommand{\vbpn}    {{\sc{vbp}}}
\newcommand{\resbosn}    {{\sc{resbos}}}
\newcommand{\powhegn}    {{\sc{powheg}}}
\newcommand{\dyradn}    {{\sc{dyrad}}}

\newcommand{\lumi}      {1~$\rm fb^{-1}$}
\newcommand{\result}    {X.X}
\newcommand{\erstat}    {^{+X.X}_{-X.X}}
\newcommand{\ersyspos}  {+X.X}
\newcommand{\ersysneg}  {-X.X}
\newcommand{\ersys}     {^{\ersyspos}_{\ersysneg}}
\newcommand{\erlumi}    {X.X}
\newcommand{\ljets}     {\ensuremath{\ell}+\rm{jets}}
\newcommand{\eplus}     {\ensuremath{e}+jets}
\newcommand{\muplus}    {\ensuremath{\mu}+jets}
\newcommand{\ltau}      {\ensuremath{\tau \ell}}
\newcommand{\etau}      {\ensuremath{\tau e}}
\newcommand{\mutau}     {\ensuremath{\tau \mu}}

\newcommand{\rs}        {XX.X}
\newcommand{\rserr}     {^{+X.X}_{-X.X}}
\newcommand{\rsigmaell}    {\ensuremath{R_{\ell\ell/\ell j}}}
\newcommand{\rsigmatau}    {\ensuremath{R_{\ltau/\ell \ell\text{-}\ell j }}}


\hspace{5.2in}

\title{First Measurement of the Angular Coefficients of  Drell-Yan $e^{+}e^{-}$ pairs in the $Z$ Mass Region from 
$p\bar{p}$ Collisions at $\sqrt{s}$=1.96 TeV}

\affiliation{Institute of Physics, Academia Sinica, Taipei, Taiwan 11529, Republic of China} 
\affiliation{Argonne National Laboratory, Argonne, Illinois 60439, USA} 
\affiliation{University of Athens, 157 71 Athens, Greece} 
\affiliation{Institut de Fisica d'Altes Energies, ICREA, Universitat Autonoma de Barcelona, E-08193, Bellaterra (Barcelona), Spain} 
\affiliation{Baylor University, Waco, Texas 76798, USA} 
\affiliation{Istituto Nazionale di Fisica Nucleare Bologna, $^{aa}$University of Bologna, I-40127 Bologna, Italy} 
\affiliation{University of California, Davis, Davis, California 95616, USA} 
\affiliation{University of California, Los Angeles, Los Angeles, California 90024, USA} 
\affiliation{Instituto de Fisica de Cantabria, CSIC-University of Cantabria, 39005 Santander, Spain} 
\affiliation{Carnegie Mellon University, Pittsburgh, Pennsylvania 15213, USA} 
\affiliation{Enrico Fermi Institute, University of Chicago, Chicago, Illinois 60637, USA}
\affiliation{Comenius University, 842 48 Bratislava, Slovakia; Institute of Experimental Physics, 040 01 Kosice, Slovakia} 
\affiliation{Joint Institute for Nuclear Research, RU-141980 Dubna, Russia} 
\affiliation{Duke University, Durham, North Carolina 27708, USA} 
\affiliation{Fermi National Accelerator Laboratory, Batavia, Illinois 60510, USA} 
\affiliation{University of Florida, Gainesville, Florida 32611, USA} 
\affiliation{Laboratori Nazionali di Frascati, Istituto Nazionale di Fisica Nucleare, I-00044 Frascati, Italy} 
\affiliation{University of Geneva, CH-1211 Geneva 4, Switzerland} 
\affiliation{Glasgow University, Glasgow G12 8QQ, United Kingdom} 
\affiliation{Harvard University, Cambridge, Massachusetts 02138, USA} 
\affiliation{Division of High Energy Physics, Department of Physics, University of Helsinki and Helsinki Institute of Physics, FIN-00014, Helsinki, Finland} 
\affiliation{University of Illinois, Urbana, Illinois 61801, USA} 
\affiliation{The Johns Hopkins University, Baltimore, Maryland 21218, USA} 
\affiliation{Institut f\"{u}r Experimentelle Kernphysik, Karlsruhe Institute of Technology, D-76131 Karlsruhe, Germany} 
\affiliation{Center for High Energy Physics: Kyungpook National University, Daegu 702-701, Korea; Seoul National University, Seoul 151-742, Korea; Sungkyunkwan University, Suwon 440-746, Korea; Korea Institute of Science and Technology Information, Daejeon 305-806, Korea; Chonnam National University, Gwangju 500-757, Korea; Chonbuk National University, Jeonju 561-756, Korea} 
\affiliation{Ernest Orlando Lawrence Berkeley National Laboratory, Berkeley, California 94720, USA} 
\affiliation{University of Liverpool, Liverpool L69 7ZE, United Kingdom} 
\affiliation{University College London, London WC1E 6BT, United Kingdom} 
\affiliation{Centro de Investigaciones Energeticas Medioambientales y Tecnologicas, E-28040 Madrid, Spain} 
\affiliation{Massachusetts Institute of Technology, Cambridge, Massachusetts 02139, USA} 
\affiliation{Institute of Particle Physics: McGill University, Montr\'{e}al, Qu\'{e}bec, Canada H3A~2T8; Simon Fraser University, Burnaby, British Columbia, Canada V5A~1S6; University of Toronto, Toronto, Ontario, Canada M5S~1A7; and TRIUMF, Vancouver, British Columbia, Canada V6T~2A3} 
\affiliation{University of Michigan, Ann Arbor, Michigan 48109, USA} 
\affiliation{Michigan State University, East Lansing, Michigan 48824, USA}
\affiliation{Institution for Theoretical and Experimental Physics, ITEP, Moscow 117259, Russia}
\affiliation{University of New Mexico, Albuquerque, New Mexico 87131, USA} 
\affiliation{Northwestern University, Evanston, Illinois 60208, USA} 
\affiliation{The Ohio State University, Columbus, Ohio 43210, USA} 
\affiliation{Okayama University, Okayama 700-8530, Japan} 
\affiliation{Osaka City University, Osaka 588, Japan} 
\affiliation{University of Oxford, Oxford OX1 3RH, United Kingdom} 
\affiliation{Istituto Nazionale di Fisica Nucleare, Sezione di Padova-Trento, $^{bb}$University of Padova, I-35131 Padova, Italy} 
\affiliation{LPNHE, Universite Pierre et Marie Curie/IN2P3-CNRS, UMR7585, Paris, F-75252 France} 
\affiliation{University of Pennsylvania, Philadelphia, Pennsylvania 19104, USA}
\affiliation{Istituto Nazionale di Fisica Nucleare Pisa, $^{cc}$University of Pisa, $^{dd}$University of Siena and $^{ee}$Scuola Normale Superiore, I-56127 Pisa, Italy} 
\affiliation{University of Pittsburgh, Pittsburgh, Pennsylvania 15260, USA} 
\affiliation{Purdue University, West Lafayette, Indiana 47907, USA} 
\affiliation{University of Rochester, Rochester, New York 14627, USA} 
\affiliation{The Rockefeller University, New York, New York 10065, USA} 
\affiliation{Istituto Nazionale di Fisica Nucleare, Sezione di Roma 1, $^{ff}$Sapienza Universit\`{a} di Roma, I-00185 Roma, Italy} 

\affiliation{Rutgers University, Piscataway, New Jersey 08855, USA} 
\affiliation{Texas A\&M University, College Station, Texas 77843, USA} 
\affiliation{Istituto Nazionale di Fisica Nucleare Trieste/Udine, I-34100 Trieste, $^{gg}$University of Trieste/Udine, I-33100 Udine, Italy} 
\affiliation{University of Tsukuba, Tsukuba, Ibaraki 305, Japan} 
\affiliation{Tufts University, Medford, Massachusetts 02155, USA} 
\affiliation{University of Virginia, Charlottesville, VA  22906, USA}
\affiliation{Waseda University, Tokyo 169, Japan} 
\affiliation{Wayne State University, Detroit, Michigan 48201, USA} 
\affiliation{University of Wisconsin, Madison, Wisconsin 53706, USA} 
\affiliation{Yale University, New Haven, Connecticut 06520, USA} 
\author{T.~Aaltonen}
\affiliation{Division of High Energy Physics, Department of Physics, University of Helsinki and Helsinki Institute of Physics, FIN-00014, Helsinki, Finland}
\author{B.~\'{A}lvarez~Gonz\'{a}lez$^w$}
\affiliation{Instituto de Fisica de Cantabria, CSIC-University of Cantabria, 39005 Santander, Spain}
\author{S.~Amerio}
\affiliation{Istituto Nazionale di Fisica Nucleare, Sezione di Padova-Trento, $^{bb}$University of Padova, I-35131 Padova, Italy} 

\author{D.~Amidei}
\affiliation{University of Michigan, Ann Arbor, Michigan 48109, USA}
\author{A.~Anastassov}
\affiliation{Northwestern University, Evanston, Illinois 60208, USA}
\author{A.~Annovi}
\affiliation{Laboratori Nazionali di Frascati, Istituto Nazionale di Fisica Nucleare, I-00044 Frascati, Italy}
\author{J.~Antos}
\affiliation{Comenius University, 842 48 Bratislava, Slovakia; Institute of Experimental Physics, 040 01 Kosice, Slovakia}
\author{G.~Apollinari}
\affiliation{Fermi National Accelerator Laboratory, Batavia, Illinois 60510, USA}
\author{J.A.~Appel}
\affiliation{Fermi National Accelerator Laboratory, Batavia, Illinois 60510, USA}
\author{A.~Apresyan}
\affiliation{Purdue University, West Lafayette, Indiana 47907, USA}
% ...
\collaboration{CDF Collaboration\footnote{With visitors from $^a$University of Massachusetts Amherst, Amherst, Massachusetts 01003,
$^b$Istituto Nazionale di Fisica Nucleare, Sezione di Cagliari, 09042 Monserrato (Cagliari), Italy,
$^c$University of California Irvine, Irvine, CA  92697, 
$^d$University of California Santa Barbara, Santa Barbara, CA 93106
$^e$University of California Santa Cruz, Santa Cruz, CA  95064,
$^f$CERN,CH-1211 Geneva, Switzerland,
$^g$Cornell University, Ithaca, NY  14853, 
$^h$University of Cyprus, Nicosia CY-1678, Cyprus, 
$^i$University College Dublin, Dublin 4, Ireland,
$^j$University of Fukui, Fukui City, Fukui Prefecture, Japan 910-0017,
$^k$Universidad Iberoamericana, Mexico D.F., Mexico,
$^l$Iowa State University, Ames, IA  50011,
$^m$University of Iowa, Iowa City, IA  52242,
$^n$Kinki University, Higashi-Osaka City, Japan 577-8502,
$^o$Kansas State University, Manhattan, KS 66506,
$^p$University of Manchester, Manchester M13 9PL, England,
$^q$Queen Mary, University of London, London, E1 4NS, England,
$^r$University of Melbourne, Victoria 3010, Australia,
$^s$Muons, Inc., Batavia, IL 60510,
$^t$Nagasaki Institute of Applied Science, Nagasaki, Japan, 
$^u$National Research Nuclear University, Moscow, Russia,
$^v$University of Notre Dame, Notre Dame, IN 46556,
$^w$Universidad de Oviedo, E-33007 Oviedo, Spain, 
$^x$Texas Tech University, Lubbock, TX  79609, 
$^y$Universidad Tecnica Federico Santa Maria, 110v Valparaiso, Chile,
$^z$Yarmouk University, Irbid 211-63, Jordan,
$^{hh}$On leave from J.~Stefan Institute, Ljubljana, Slovenia, 
}}
\noaffiliation


\date{May. 11, 2011  }

\begin{abstract}
We report on the first measurement of the angular distributions of final state electrons
 in $p\bar{p}\to \gamma^{*}/Z\to e^{+}e^{-}+X$ events produced in the $Z$ boson mass region
 at $\sqrt{s}=1.96$~TeV. The data sample collected by the CDF II detector for this result corresponds to 
 $2.1$~fb$^{-1}$ of integrated luminosity.
 The angular distributions are studied as a function of the transverse momentum of the
electron-positron pair and show good agreement with the Lam-Tung relation,
consistent with a spin-1 description of the gluon, and demonstrate that at high
values of the transverse momentum, $Z$  bosons are produced via quark anti-quark
annihilation and quark-gluon Compton processes.
\end{abstract}

\maketitle 


We report on a study of the  angular distributions of final state electrons
 in $p\bar{p}\to \gamma^{*}/Z\to e^{+}e^{-}+X$ Drell-Yan events to probe $Z$-boson production mechanisms. 
In quantum chromodynamics (QCD) at order $\alpha_s$
 this occurs either through the annihilation
process with  a gluon (G)  in the final state ($q\bar{q}\to \gamma^{*}/Z~G$),
or via  the Compton process with a quark in the final state ($qG \to \gamma^{*}/Z~q$).
The emission of final state $q/G$ gives $\gamma^{*}/Z$ transverse momentum \cite{cdf_coord}
 (we define the production $P_T=P_T(\gamma^{*}/Z)=P_T(e^+e^-)$ before final state radiation). 

The general expression for the angular distribution \cite{mirkes2}  is described by 
the polar ($\theta$) and azimuthal ($\phi$) angles of the decay-electron in the Collins-Soper (CS) frame \cite{collins}.
When integrated over  $\cos\theta$ or $\phi$, respectively, the decay-electron angular distribution is described by:
\begin{eqnarray}
\label{etheta}
\frac{d{\sigma}}{d\cos{\theta}} \propto (1+\cos^2{\theta})
+ \frac{1}{2}A_0(1-3\cos^2{\theta}) + A_4\cos{\theta}\\
\label{ephi}
\frac{d{\sigma}}{d\phi} \propto 1+\beta_3\cos{\phi} + \beta_2 \cos 2{\phi} + \beta_7 \sin{\phi} + \beta_5 \sin 2{\phi}
\end{eqnarray}
 where $\beta_3 = 3\pi A_3/16$, $\beta_2 = A_2/4$, $\beta_7 = 3\pi A_7/16$, $\beta_5 = A_5/4$.
The $A_{0}$ and $A_{4}$ are extracted from  Eq. \ref{etheta}, and $A_{2}$ and $A_{3}$ are extracted from Eq. \ref{ephi}, 
while $A_{5}$ and $A_{7}$ are expected to be zero \cite{mirkes2}.

 Perturbative QCD (pQCD) makes definite predictions for the angular coefficients $A_{0,2,3,4}$
 ($A_0$ and $A_2$ are the same for $\gamma^{*}$ or $Z$ exchange, and $A_3$ and $A_4$ originate from the $\gamma^{*}/Z$ interference).
 For the $q\bar{q}\to \gamma^{*}/Z~G$ annihilation process 
pQCD at order $\alpha_s$ predicts that  the angular coefficients 
$A_0$ and $A_2$  are equal \cite{qq,bnl, berger,bodek} 
   and can be analytically described by  
$ A_0^{q\bar{q}} =  A_2^{q\bar{q}}= P_T^2/(M_{e^{+}e^{-}}^2+P_T^2)$ (Eq. 3).
At higher order, there are small deviations from the above
expression (Eq. 3) which depend on  PDFs and  dilepton rapidity ($y$) \cite{cdf_coord}.

For the $qG \to \gamma^{*}/Z~q$ Compton process, $A_0$ and $A_2$ 
depend on parton distribution functions (PDFs) and $y$.  However, in
pQCD at order $\alpha_s$, when averaged over $y$,  $A_0$ and $A_2$ are approximately 
described \cite{dyno1,gq} by $A_0^{qG} = A_2^{qG} \approx 5P_T^2/(M_{e^{+}e^{-}} ^2+5P_T^2)$ (Eq. 4).

At order $\alpha_s$, the Lam-Tung relation ($A_0= A_2$) \cite{LT} is valid for both $q\bar{q}$ and $qG$ processes \cite{bnl}. 
Fixed-order pQCD calculations at
 order $\alpha_s^2$ \cite{mirkes2}, as well as QCD  resummation
 calculations to all orders \cite{berger}, indicate that  violations of the Lam-Tung relation are small. 
The Lam-Tung relation is only valid for vector (spin-1) gluons. 
It is badly broken for  scalar (spin-0) gluons \cite{scalar}.
Therefore,  confirmation of the Lam-Tung relation is a fundamental test of the vector
gluon nature of  QCD and is equivalent to a measurement of the spin of the gluon.
A previous determination of the gluon spin was made from a study of 3-jet events
 ($ e^+e^-\to q\bar{q}~G$) in $e^+e^-$ annihilation \cite{tasso}.

To date, the Lam-Tung relation has been tested only  
at fixed-target experiments using samples of low mass Drell-Yan dilepton
pairs at relatively low transverse momentum. In this region,  non-perturbative
higher-twist effects  can be significant \cite{boer,dyno3}.  Some  experiments report
large violations \cite{dyno1,dyno3,dyno2, dyno4}, and one experiment \cite {dyok}
is consistent with the Lam-Tung relation.  
 Here we report on the first test of the Lam-Tung relation at large dilepton
mass and high transverse momentum, where non-perturbative
higher-twist effects are expected to be negligible.

\begin{figure}
\includegraphics[width=3in,height=7.6in]{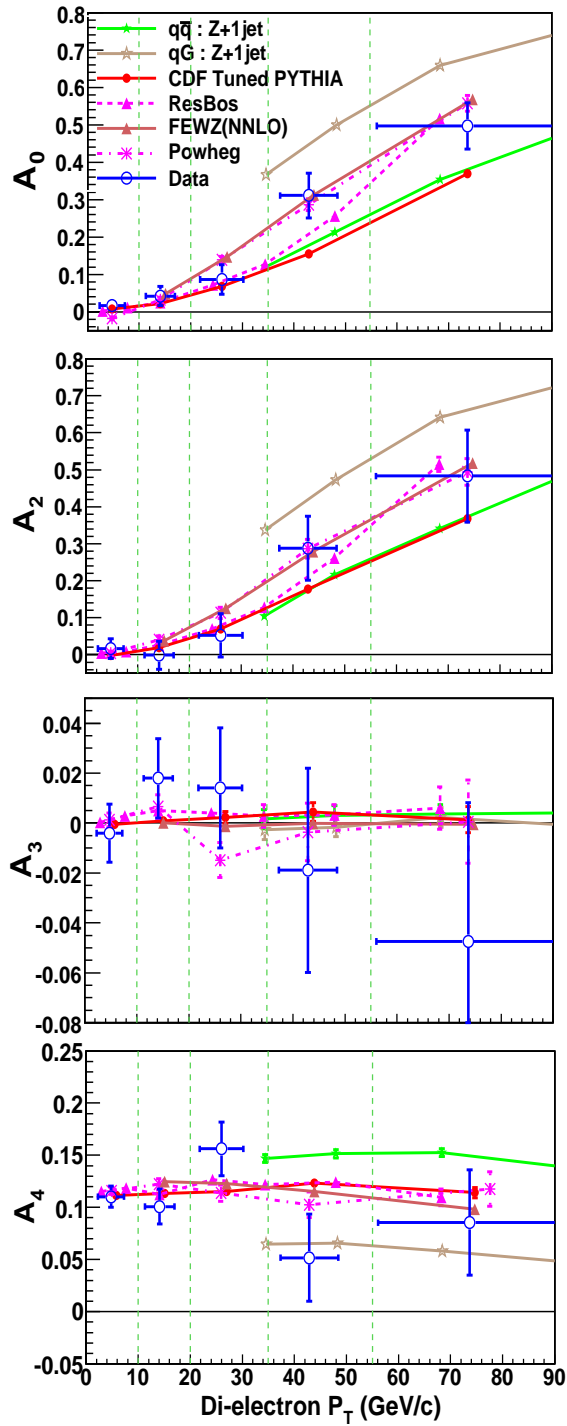}
\caption{\label{fig:3}Comparison of the measured values of $A_0$, $A_2$, $A_3$ and  $A_4$ 
 (for $66<M_{ee}<116$~GeV/c$^2$), shown with statistical and systematic uncertainties combined
 in quadrature, to theory predictions.
 The data are plotted at the mean $P_T$ of the events for each bin. The last bin corresponds to $P_T > 55$ GeV/c with no upper limit.
 The horizontal uncertainty is RMS of the transverse momenta in each bin.
  Agreement \cite{chi} is found with the predictions of   \fewz and \powheg (shown) ,  and also with \dyrad, \madgraph,  
and \pythia $~Z$ +1-jet MC (not shown).  The data  do not favor \cite{chi} the predictions of  default  \pythia, and  \vbp.
 Also shown are the pure  $q\bar{q}\to \gamma^{*}/Z~G$  annihilation diagram prediction 
  and the  $qG \to \gamma^{*}/Z~q$ Compton process prediction from the  \pythia $~Z$ +1-jet MC. 
}
\end{figure}

Fixed order pQCD calculations \cite{mirkes2} and Monte
Carlo (MC) simulations at next-to-leading order (NLO) (e.g.  \dyrad \cite{dyrad}  and \madgraph \cite{madgraph},
and  \pythia in Z+1jet mode \cite{pythia64}) indicate that  there is  a significant ($\approx$ 30\%)
contribution of the Compton process to the production of
 $\gamma^{*}/Z$  bosons 
 at the Tevatron.
Therefore, as shown in Fig. \ref{fig:3}, these calculations yield values of $A_0$ and $A_2$
 which are larger than the pure annihilation process prediction (Eq. 3).
Similar results are predicted by \powheg \cite{powheg}, a NLO MC with
additional parton showering,  and \fewz \cite{FEWZ} which is a next-to-next-to-leading order (NNLO) 
QCD calculation.

In contrast, the default, LO version of \pythia \cite{pythia}, and \vbp \cite{vbp} 
(an MC generator based on QCD resummation) predict values of  $A_0$ and $A_2$ which are close to Eq. 3 
 (which is only correct if the  $q\bar{q}$ process is dominant).
 The  \resbos \cite{resbos} MC generator,  which is also based on QCD resummation,  
predicts values of $A_0$ and $A_2$ close to Eq. 3 at low $P_T$, and larger values
 (close to the predictions of fixed order pQCD) at high $P_T$, as shown in Fig. \ref{fig:3}.
Therefore, measurements of $A_0$ and $A_2$
as a function of $P_T$ elucidate the relative contributions between
the annihilation and Compton processes.

 In this Letter, we report on the first
measurement of the angular coefficients  $A_0$, $A_2$, $A_3$ and $A_4$,
 for $p\bar{p}\to \gamma^{*}/Z\to e^{+}e^{-}+X$ events
in the $Z$ boson mass region ($66<M_{ee}<116$~GeV/c$^2$)
produced  at $\sqrt{s}=1.96$~TeV.  We also report on the first test of
the Lam-Tung relation at  high transverse momentum. 

The sample used  corresponds to an integrated luminosity of 
  $2.1$ fb$^{-1}$ collected by the CDF II Detector
 at Fermilab \cite{cdfdet} during 2004-2007. 
Charged particle directions and momenta are measured by an open-cell drift
chamber (COT), a silicon vertex detector (SVX), and an intermediate silicon layer
 in a 1.4 T magnetic field.  Projective-tower-geometry calorimeters
and outer muon detectors enclose the magnetic tracking volume.
The coverage of COT tracking in pseudorapidity is $|\eta|<1.2$ \cite{cdf_coord}. 
Reconstructed tracks are used to determine 
the $p\bar{p}$ collision point along the beam line, 
which is required to be within $z=\pm60$ cm of the center of the detector.
The energies and directions \cite{cdf_coord} of electrons, photons,
and jets are measured by two separate calorimeters:
central ($|\eta|<1.1$) and plug ($1.1<|\eta|<3.6$).
Each calorimeter  has an electromagnetic compartment with a
shower maximum detector followed by a hadronic compartment.
Three topologies of $e^{+}e^{-}$ pairs are considered: two central electrons (CC),
 one central and one plug electron (CP), and two plug electrons (PP). 
 Events with at least one electron with  high $E_{T}$ are selected online.
 Offline refined selection requires the electron to have 
  $E_{T}>25$ GeV for CC and PP events, and $E_{T}>20$ GeV for CP events 
 in the fiducial regions of the calorimeters,
 the central ($|\eta_e|<1.1$) and plug ($1.2<|\eta_e|<2.8$).
To minimize background, the  second electron candidate is required to have   $E_{T}>15$ GeV for  
  CC,  $E_{T}>25$ GeV for  PP, and $E_{T}>20$ GeV  for CP events.
The selection criteria listed above
are the same as in the related previous publication \cite{dsdy} of the $Z$ rapidity
distribution, but are augmented in this analysis with the additional
requirement that both electrons have an associated track in the SVX.
The data sample  consists of  about 140 000 events.  
The fractional contribution of the total  QCD background (2-jet events misidentified as a Drell-Yan  pairs)
 to the number of selected events is $0.3\%$. This is determined by studying 
the distribution of transverse energy in a cone surrounding the center of the electromagnetic
 cluster in the calorimeter.
 The  total  background from electroweak ($WW$, $WZ$, $W$+jets, 
 and $Z\to\tau^{+}\tau^{-}$) and $t\bar{t}$ processes is estimated from simulation to be $0.2\%$.

\begin{figure}
\includegraphics[width=3.0in,height=2.0in]{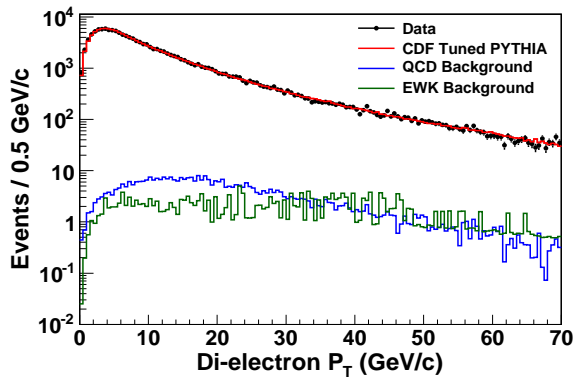}
\caption{\label{fig:1} Di-electron $P_{T}$ spectrum of data, default (CDF Tuned)  \pythia  prediction, and backgrounds (QCD and electroweak process).
The mass range corresponds to $66 < M_{ee} < 116$ GeV/c$^{2}$.}
\end{figure}

\begin{figure}
\includegraphics[width=3.0in]{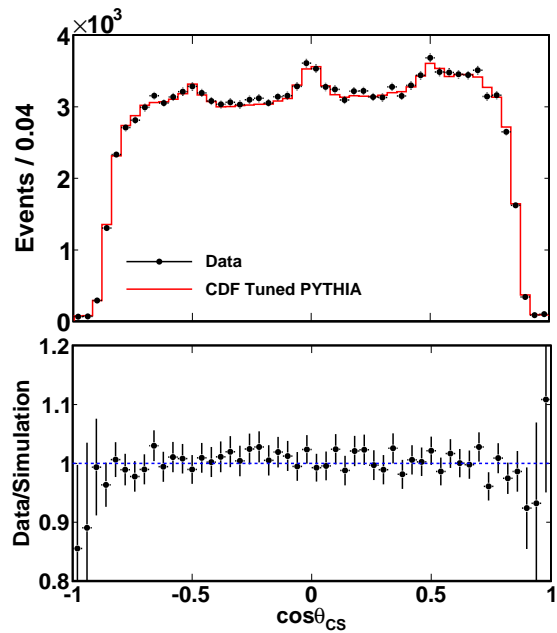}
\caption{\label{fig:2} The $\cos\theta$ distribution of data and default (CDF Tuned) \pythia prediction.}
\end{figure}
%

%

The effect of the acceptance on the angular
distributions is modeled using the  \pythia MC generator \cite{pythia}
 combined with a {\sc geant} \cite{geant} simulation of the CDF detector.
The  \pythia generator includes a LO QCD interaction ($q\bar{q} \to \gamma^{*}/Z$),
 initial state QCD radiation, parton shower fragmentation, the 
 $\gamma^{*}/Z \to e^{+}e^{-}$ decay, and  photon radiation from the final state. 
 The version of \pythia  used at CDF has additional ad-hoc tuning \cite{pythia} 
 (referred to as default {\sc{pythia}}) in order to accurately represent the
 $\gamma^{*}/Z$  boson transverse momentum distribution measured in data. 
 Further tuning was introduced in order to ensure that the MC simulation correctly
 described the rapidity, as well as the correlations between rapidity
 and transverse momentum that are observed in the data.
  To reconstruct the simulated events in the same way as data, the calorimeter energy scale, 
  resolutions, and selection efficiencies used in the detector simulation are tuned \cite{dsdy} using data.
 Figure \ref{fig:1} shows the di-electron $P_{T}$ spectrum for data, the default \pythia prediction, and the backgrounds.
 There is good agreement between data and \pythia prediction.
Figure \ref{fig:2} shows the $\cos\theta$ distribution for data and the default \pythia prediction and its ratio.

The analysis is performed in five bins of transverse momentum as shown
 in Table~\ref{A:coeff}.  For each transverse momentum range,  data
 and MC simulated events are binned in $\cos\theta$ and $\phi$.
  The MC events are re-weighted to generate the expected angular distributions ($\cos\theta$ and $\phi$) for a range
 of values of $A_0$ and $A_4$, and $A_2$ and $A_3$, respectively.
The angular distributions from the re-weighted MC events are compared to the data in the reconstructed level
 and the angular coefficients which give a maximum log-likelihood value are determined as the best coefficients to describe the data.  
The $A_0$ and $A_4$ are determined by the comparison of the data to MC distributions in $\cos\theta$ and
 the $A_2$ and $A_3$ are determined in $\phi$. 
The normalization factor of the data to MC events is also included as one of fit parameters.
 The results are  shown in  Fig. \ref{fig:3}, and in  Table~\ref{A:coeff} with statistical 
 and systematic uncertainties. The correlation between extracted values of $A_0$ and $A_2$, $A_3$ and $A_4$ is negligible.
 The systematic uncertainties originating from backgrounds, electron identification efficiency, SVX tracking efficiency,
 boson $P_{T}$ and rapidity modeling, and modeling of detector material are considered. The dominant source is the background estimate.
 Most of systematic uncertainties are discussed in reference \cite{dsdy} and
 the effect of these uncertainties on the shape of the angular distribution is small.

The data are in good agreement with the Lam-Tung relation $A_0-A_2=0$, which is expected in QCD with vector gluons. 
The  values of $A_0-A_2$ for the five $P_T$ bins are $0.00 \pm 0.03$, $0.04 \pm 0.05$,
$0.03 \pm 0.07$, $0.02 \pm 0.11$, and $0.01 \pm 0.14$ (statistical and systematic uncertainties combined),
 which average to  $ \langle A_0-A_2 \rangle $=$0.02 \pm 0.02$.
At low  $P_T$ the measured values of  $A_0$ and $A_2$
are  well described by the $q\bar{q}\to \gamma^{*}/Z~G$ annihilation function (Eq. 3).  
 At high $P_T$ the larger values show that both the annihilation and Compton processes contribute to the cross section \cite{chi}. 
 Our results are in agreement \cite{chi}  with fixed-order perturbation theory calculations including 
  \dyrad \cite{dyrad},  \madgraph \cite{madgraph}, \pythia Z+1 jet~\cite{pythia64},  
 \powheg \cite{powheg},  and  \fewz \cite{FEWZ} (all of these give similar predictions). 
  We find that the values of  $A_3$ and $A_4$  are in agreement  with the predictions of all  models
  ($A_4$ is calculated with $\sin^2 \theta_{\rm W} = 0.232$).

   In summary, we present the first measurement
  of the angular coefficients in the production of  $\gamma^{*}/Z$ bosons at large
  transverse momenta, and the first test
  of the Lam-Tung relation at  high transverse momentum.
    We find good agreement with the predictions of
  QCD  fixed-order perturbation theory, and with  the  Lam-Tung relation $A_0=A_2$. 
 The measurements presented here are statistically limited. An analysis with
 larger samples in both muon and electron channels is currently
under way.  A comparison of these results with future measurements
at the  LHC would provide additional tests of production mechanisms since  the contribution of the Compton process ($qG \to \gamma^{*}/Z~q$) at the LHC is expected to be larger.

\begin{table}[ht]
\caption{\label{A:coeff} The measured angular coefficients (measured value $\pm$ stat. error $\pm$ syst. error). 
The mean $P_T$ of the events in the five bins are 4.8, 14.1, 26.0, 42.9, and 73.7 GeV/c, respectively. }
\begin{tabular}{|c|c|c|} \hline \hline
$P_{\rm T}$ bin &  $A_0$ ($\times 10^{-1}$)  &  $A_2$ ($\times 10^{-1}$)  \\ \hline
0--10 
    &   $0.17 \pm 0.14 \pm 0.07$   &  $0.16 \pm 0.26 \pm 0.06$    \\ 
10--20
    &   $0.42 \pm 0.25 \pm 0.07$   &  $-0.01 \pm 0.35 \pm 0.16$   \\ 
20--35
    &   $0.86 \pm 0.39 \pm 0.08$   &  $0.52 \pm 0.51 \pm 0.29$   \\ 
35--55
    &   $3.11 \pm 0.59 \pm 0.10$   &  $2.88 \pm 0.84 \pm 0.19$  \\ 
$>$ 55
    &  $4.97 \pm 0.61 \pm 0.10$    &  $4.83 \pm 1.24 \pm 0.02$   \\\hline \hline
\end{tabular}
\begin{tabular}{|c|c|c|} \hline \hline
$P_{\rm T}$ bin &  $A_3$ ($\times 10^{-1}$)  &  $A_4$ ($\times 10^{-1}$) \\ \hline 
0--10 
     & $-0.04 \pm 0.12 \pm 0.01$  &   $1.10 \pm 0.10 \pm 0.01$    \\ 
10--20
     & $0.18 \pm 0.16 \pm 0.01$   &  $1.01 \pm 0.17 \pm 0.01$ \\ 
20--35
     & $0.14 \pm 0.24 \pm 0.01$   &  $1.56 \pm 0.26 \pm 0.01$ \\ 
35--55
     & $-0.19 \pm 0.41 \pm 0.04$   &  $0.52 \pm 0.42 \pm 0.03$ \\ 
$>$ 55
     & $-0.47 \pm 0.56 \pm 0.02$   & $0.85 \pm 0.50 \pm 0.05$   \\\hline \hline
\end{tabular}
\end{table}

We thank the Fermilab staff and the technical staffs of the participating institutions for their vital contributions. This work was supported by the U.S. Department of Energy and National Science Foundation; the Italian Istituto Nazionale di Fisica Nucleare; the Ministry of Education, Culture, Sports, Science and Technology of Japan; the Natural Sciences and Engineering Research Council of Canada; the National Science Council of the Republic of China; the Swiss National Science Foundation; the A.P. Sloan Foundation; the Bundesministerium f\"ur Bildung und Forschung, Germany; the Korean World Class University Program, the National Research Foundation of Korea; the Science and Technology Facilities Council and the Royal Society, UK; the Institut National de Physique Nucleaire et Physique des Particules/CNRS; the Russian Foundation for Basic Research; the Ministerio de Ciencia e Innovaci\'{o}n, and Programa Consolider-Ingenio 2010, Spain; the Slovak R\&D Agency; and the Academy of Finland.


\begin{thebibliography}{99}
\bibitem{cdf_coord}
In the CDF detector frame, the positive $z$ axis is define by the proton beam direction. 
The pseudorapidity is $\eta=-\ln\tan(\theta/2)$. For an $e^{+}e^{-}$ pair
 $P_{T}=P\sin\theta$,  $E_{T}=E\sin\theta$,  where $\theta$ is the polar
 angle between the particle direction and the $z$ axis. For an $e^{+}e^{-}$ pair, 
  $y=\frac{1}{2}\ln\frac{E+P_{z}}{E-P_{z}}$, where
$P$ and $P_{z}$ are the magnitude and $z$-component of the momentum,
and $E$ and $M_{e^+e^-}$ are  the energy and mass.
\bibitem{collins} J. C. Collins and D. E. Soper, Phys. Rev. D {\bf16}, 2219 (1977). The CS frame is the center of mass of the $\gamma^{*}/Z$, where the $z$ axis is defined as the bisector of the $p$ and  $\bar{p}$ beams. The polar angle is defined as the angle of
the positive lepton with respect to the proton direction.
\bibitem{chi}  For  the four highest $P_T$ bins the $\chi^2$  values for $A_0$ are
11.7, 9.4, 5.6 and 3.7 for default  \pythia ($q\bar{q}$ and t channel $qG$), \vbpn, \resbosn, and  \fewz ($q\bar{q}$+$qG$),
respectively.
\end{thebibliography}
\end{document}